\documentclass[12pt]{iopart}

\usepackage{graphicx}
\usepackage{color}
\usepackage{float}
\usepackage{iopams}
\usepackage{multirow}

\newcommand{\old}[1]{{\rule{0cm}{0cm}}}

\renewcommand{\vec}[1]{{\mathbf #1}}

\newcommand{\la}{\langle}
\newcommand{\ra}{\rangle}

\newcommand{\vnabla}{\boldsymbol{\mathbf\nabla}}

\newcommand{\vPi}{\boldsymbol{\mathbf\Pi}}
\newcommand{\vsigma}{\boldsymbol{\mathbf\sigma}}

\newcommand{\bbM}{\boldsymbol{\mathbb M}}

\newcommand{\bj}{\boldsymbol{\mathbf j}}
\newcommand{\bk}{\boldsymbol{\mathbf k}}
\newcommand{\bkp}{\boldsymbol{\mathbf k^\prime}}

\renewcommand{\br}{\boldsymbol{\mathbf r}} 
\renewcommand{\bs}{\boldsymbol{\mathbf s}}  
\newcommand{\bF}{\boldsymbol{\mathbf F}}

\newcommand{\bR}{\boldsymbol{\mathbf R}}
\newcommand{\bS}{\boldsymbol{\mathbf S}}
\newcommand{\bT}{\boldsymbol{\mathbf T}}

\newcommand{\hPsig}{\hat{P}_\sigma}
\newcommand{\dd}{\delta(\br)}

\newcommand{\occ}{\frac{(2j+1)}{4 \pi}}

\newcommand{\text}{\mathrm}

\newcommand{\CE}{\mathcal{E}}

\renewcommand{\etal}{\emph{et al.}} 
\newcommand{\nn}{\nonumber}
\newcommand{\noi}{\noindent}
\newcommand{\q}{\quad}

\newcommand{\bc}{\begin{center}}
\newcommand{\ec}{\end{center}}
\newcommand{\be}{\begin{equation}}
\newcommand{\ee}{\end{equation}}
\newcommand{\bqr}{\begin{eqnarray}}
\newcommand{\eqr}{\end{eqnarray}}
\renewcommand{\bi}{\begin{itemize}}   
\newcommand{\ei}{\end{itemize}}
\newcommand{\bwt}{ }
\newcommand{\ewt}{ }
\newcommand{\bsub}{\numparts}    
\newcommand{\esub}{\endnumparts}     
\newcommand{\tfrac}{\frac}

\newcommand{\citeqdot}[1]{Eq.~(\ref{#1})}

\newcommand{\citeAppendix}[1]{\ref{#1}}

\newcommand{\salpha}{\sum_{\{\alpha \}} \, \occ}

\newcommand{\wfusqalpha}{\left[ u_{\{\alpha \}} (r) \right]^2}

\newcommand{\wfupsqalpha}{\left[ u^{\prime}_{\{\alpha \}} (r) \right]^2}

\begin{document}

\title{Tools for incorporating a D-wave contribution in Skyrme energy density functionals}

\author{P. Becker, D. Davesne, J. Meyer}
\address{Universit{\'e} de Lyon, F-69622 Lyon, France, \\
Universit\'e Lyon 1, Villeurbanne;  CNRS/IN2P3, UMR5822, Institut de Physique Nucl\'eaire de Lyon}

\author{A. Pastore}
\address{Institut d'Astronomie et d'Astrophysique, CP 226, Universit\'e Libre de Bruxelles, B-1050 Bruxelles, Belgium}

\author{J. Navarro}
\address{IFIC (CSIC-Universidad de Valencia), Apartado Postal 22085, E-46.071-Valencia, Spain}

\date{\today}

\begin{abstract}
The possibility of adding a D-wave term to the standard Skyrme effective interaction has been widely considered in the past. Such a term has been shown to appear in the next-to-next-to-leading order of the Skyrme pseudo-potential. 
The aim of the present article is to provide the necessary tools to incorporate this term in a fitting procedure: first, a mean-field equation written in spherical symmetry in order to describe spherical nuclei and second, the response function to detect unphysical instabilities. With these tools it will be possible to build a new fitting procedure to determine the coupling constants of the new functional.
\end{abstract}



\date{\today}


\maketitle


\section{Introduction}
\label{sect:intro}


The nuclear energy density functional (NEDF) method is a tool of choice~\cite{RMP} 
to treat the many-body problem in the medium-heavy region of the nuclear mass chart. 
The form of the energy functional to be used is not known a priori and there exists thus a great deal of versatility regarding its parameterization. In practice, two families of non-relativistic NEDF parameterizations are mainly used, 
the one derived from the local Skyrme~\cite{Sky56a,Sky59b} interaction and the one obtained from the non-local Gogny~\cite{Dec80a} one. 

The very first applications of the Skyrme pseudo-potential for calculating ground state properties of even-even nuclei was done by Brink and Vautherin~\cite{VB72} by means of a self-consistent Hartree-Fock (HF) calculation~\cite{Rin80aB}.
In this case, the authors slightly modified the form of the original Skyrme interaction~\cite{Sky56a,Sky59b} to simplify the resulting HF calculations.
 In particular the tensor and higher order momentum terms ($i.e.$ D-wave) were removed, and the contact three-body term was replaced with a density-dependent two-body term.
Hereafter, the Skyrme pseudo-potential assumed the $standard$ form that is widely used nowadays~\cite{Cha97a,Cha98a}, which can reproduce with a reasonable accuracy several observables of both finite nuclei and infinite nuclear matter.
From this pseudo-potential it is possible to derive a functional~\cite{per04}, which is very useful in practical calculations. The relation between functional and pseudo-potential is imposed by some specific relations between the coupling constants of the functional itself~\cite{rai11}. 

Recently, a large scientific collaboration, named UNEDF~\cite{ber12,bog13,link}, has studied the optimization procedure used to determine the coupling constants of the Skyrme functional~\cite{kor10,kor12}. The authors have focused on the time even part of the standard Skyrme fuctional~\cite{per04}  to discern whether or not  its spectroscopic qualities could be improved  using the state of the art optimization procedure. In the latest article of the UNEDF collaboration~\cite{kor14}, the authors concluded that it is not possible to improve the agreement between experimental observables and theoretical calculations based on Skyrme functionals at single-reference level~\cite{lac09}. It is thus mandatory to go $beyond$ the  $standard$ form of Skyrme functionals. Two possible ways can be identified: (i) following the spirit of  the self-consistent mean field theory, where the major ingredient is an effective pseudo-potential and where the correlations beyond mean-field are added afterwards~\cite{RMP}, (ii) using the NEDF theory, where the building block is the functional which includes all correlations~\cite{gor13}.
Concerning the first approach, it is worth mentioning the recent works concerning the exploration of other additional tensor terms~\cite{Les07a} or a general three body contact term~\cite{sad13} into the Skyrme pseudo-potential. In the present article, we continue this exploration of extra terms by investigating the role of gradient terms in the central part of the pseudo-potential.

The study of higher-order terms has been systematically performed in Refs.~\cite{rai11,car08}. The idea behind the inclusion of  higher derivative terms is to mimic the presence of a finite range in the nuclear force. In its original article Skyrme introduced for the very first time this concept~\cite{Sky56a}, but he stopped the development at second order, although in the same article he mentioned the possible importance of fourth order terms.
To quantify the quality of this approximation, we refer to a very recent study~\cite{car11} done by  Carlsson and collaborators  within the context of Density Matrix Expansion (DME). It has been shown that the inclusion of 4th order terms improves the agreement among the calculations done with the complete finite range pseudo-potential ($i.e.$ Gogny)  and the DME calculations by one order of magnitude going from an average difference of $\approx10$ MeV at $2nd$ order to $\approx1$MeV at $4th$ order.  The inclusion of 6th order term improves further the agreement, but the relative gain is not so important as in the previous case. 

Following this motivation, we have investigated in ref.~\cite{dav13} the explicit form up to the fourth-order of the Skyrme pseudo-potential in Cartesian basis, compatible with Galilean and gauge invariance. 
It is important to notice that the original extra term suggested by Skyrme~\cite{Sky56a} and called D-wave,  does not satisfy the gauge-invariance symmetry~\cite{dob96} and the resulting pseudo-potential violates the continuity equation~\cite{rai11b}. Moreover it does not contribute to some important properties of infinite nuclear matter as the equation of state (EoS).
In contrast, the new terms deduced in~\cite{dav13} are gauge invariant by construction and they do give non-zero contribution to the EoS of infinite nuclear matter~\cite{dav14b}. These terms are actually a mix of a S and D partial wave.
In this work, we continue our previous investigation by giving two important tools that are required to incorporate the 4th order terms into a fitting procedure to fix its coefficients. In particular we have noticed that the major modification come from the D-wave coupling, while the S-wave term at $4th$ order does not introduce difficult changes. For such a reason we will usually speak about D-wave terms only although to respect gauge invariance we have been obliged to consider also 4th order terms in S-wave.

The article is organized as follows: in Sec.~\ref{sect:skyrme}, we investigate the properties of the functional derived from the 4-th order pseudo-potential, in particular we introduce all the necessary fields in spherical symmetry to be injected into a Schr\"odinger equation to solve HF equations.
In Sec.~\ref{LR:skyrme}, we derive the formalism of the Linear Response (LR) theory for these extra terms and present in particular the general expression of the inverse-eergy-weighted sum rule, which is the tool of choice to detect possible instabilities. We present our main conclusions in Sec.~\ref{conclusions}.


\section{Formalism}
\label{sect:skyrme}


\subsection{The energy density functional}

%

The standard parameterization of the local Skyrme NEDF kernel reads as the sum 
of a kinetic term, the Skyrme potential term
that models the contribution from the strong force 
in the \emph{particle-hole} channel, a pairing term, the Coulomb term 
(calculated using the Slater approximation~\cite{Sla51}) and a correction 
term that approximately removes the excitation energy from the spurious motion 
of the localized center of mass
\be
\label{eq:efu:complete}
\CE = \CE_{\text{kin}}
    + \CE_{\text{Sk}}
    + \CE_{\text{pair}}
    + \CE_{\text{Coul}}
    + \CE_{\text{cm}}     \,.
\ee
The Skyrme potential energy, $ \CE_{\text{Sk}}$, can be parametrized 
directly~\cite{per04} or  derived as the average value of an  
effective interaction in a Slater determinant state. 
The latter approach induces interrelations between the coupling constants 
entering the NEDF kernel and thus reduces the number of free parameters as compared 
to the former~\cite{rai11}.
The advantage of using a functional based on an effective interaction instead of a general one is to avoid in a simple manner spurious instabilities in  multi-reference calculations~\cite{lac09,Lacroix2,Lacroix3,cha10}. 
If one is interested 
in deriving both $\CE_{\text{Sk}}$ and $\CE_{\text{pair}}$ from the same 
effective interaction, one must compute their average values 
in a Bogoliubov state.
%
\subsubsection{Skyrme interaction with D-wave term.}
%

The generalized Skyrme effective interaction considered in this paper reads
\be
\label{skyrmestart}
   v^{\text{Sk}} \, \equiv \, v^{\text{(0)}}
                     \, + v^{\text{(2)}}
                     \, + \, v^{(4)}  \, + \, v^{\text{LS}}    \, + \, v^{\text{T}} + \, v^{\text{3b}}    \,   \,,
\ee
where the different terms  $v^{\text{(0)}}, v^{\text{(2)}},v^{(4)}$ corresponds to the different contribution order by order to the central term~\cite{rai11,dav13,dav14b}. They read
\begin{eqnarray}
\fl & v^{(0)} =  t_0 \, ( 1 + x_0 \hPsig ) \,, \label{eq:Skyrme:0}\\
\fl & v^{(2)} =  \tfrac{1}{2} \, t_1 \, ( 1 + x_1 \hPsig )
      \big[ \bk^{\prime 2}  + \bk^2 \big]         +  t_2 \, ( 1 + x_2 \hPsig ) \, \bkp \cdot  \bk    \, , \label{eq:Skyrme:2}\\
\fl & v ^{(4)}  =   \frac{1}{4} t_1^{(4)} (1+x_1^{(4)} P_{\sigma}) \left[({\bf k}^2 + {\bf k'}^2)^2 + 4 ({\bf k'} \cdot {\bf k})^2\right] \nn\\
\fl & \quad \quad + t_2^{(4)} (1+x_2^{(4)} P_{\sigma}) ({\bf k'} \cdot {\bf k}) ({\bf k}^2 + {\bf k'}^2) \label{V-four}\,,
\end{eqnarray}
where the definition of $\br$, $\bR$, $\bk$, $\bkp$ and $\hPsig$ is standard 
and can be found in the review paper of Bender \etal~\cite{RMP}.  In terms of partial waves the 0th order contains only S wave, the 2nd order is a mixture of S and P waves, while the 4th order mixes S, P and D waves. As already briefly explained in the introduction, the major modifications comes from the D-wave coupling, while the role of the higher order S-wave is to satisfy the gauge invariance. The spin-orbit term simply reads
\begin{eqnarray}
\label{eq:Skyrme:LS}
v^\mathrm{LS} (\br) & = & i \, W_0 \, ( \vsigma_1 + \vsigma_2 ) \cdot \left[ \bk^{\prime} \times \dd \; \bk \right]\,. 
\end{eqnarray}
The inclusion of higher order terms does not affect the spin-orbit term. In fact as already discussed  in ref.~\cite{dav13}, this is the only possible gauge-invariant structure we can build. A contribution to the spin-orbit term, which respects gauge invariance, could only come from tensor terms  $v^{\text{T}}$~\cite{Les07a,dav13} which have been discarded here for simplicity. We refer to the discussion in Ref.~\cite{dav13} for more details.
Finally we have the three-body term $v^{\text{3b}}$, which has been recently investigated by Sadoudi \emph{et al.}~\cite{sad13}. In the present article we substitute it with a simple density dependent term as~\cite{VB72}
\begin{eqnarray}\label{sky:dd}
v^\mathrm{3b} (\br) & \approx &v^\mathrm{DD} (\br) =\frac{1}{6} t_3 \, ( 1 + x_3 \hPsig )\rho^{\alpha}(\mathbf{R}) \; \dd \,.
\end{eqnarray}
This corresponds to taking only the simplest term composed by three Dirac delta in the construction of the three-body potential~\cite{sad13}; for $\alpha=1$ there is in fact an equality of their respective Hartree-Fock expectation values. 
The use of an integer power for $\alpha$ (but not necessary 1) is required to avoid some of instabilities in multi-reference calculations~\cite{was12}.
The inclusion of an explicit three body term would slightly change the equations presented in this article, but this would not represent a big effort compared to the one of including the D-wave coupling. 
In this simplified version of the Skyrme pseudo-potential in Eq.~(\ref{skyrmestart}), we need to constrain 2 coefficients at 0th order, 4 at 2nd order and 4 at 4th order, 1 spin-orbit parameter and 3 coefficients for the density dependent term of Eq.~(\ref{sky:dd}). The total number of coefficients is thus 14. 
Using a real three-body term and a first order tensor term would increase the parameters to 18. Although it is a large number compared to standard Skyrme pseudo-potential~\cite{Cha97a}, it is however smaller than the number of free parameters used in many DFT based approaches~\cite{gor13,Cha08a}. It is thus possible to determine a new fitting protocol to determine these parameters.
%
%
\subsubsection{Local densities and currents.}
%

Neutron ($q=n$) and proton ($q=p$) density matrices are written 
in position$\otimes$spin$\otimes$isospin space according to
\bqr
\rho_q (\br \sigma ,\br' \sigma')
& = &   \tfrac{1}{2} \, \rho_q (\br,\br') \delta_{\sigma \sigma'}
      + \tfrac{1}{2} \, \bs_q (\br,\br')
        \cdot \la \sigma' | \hat{\vsigma} | \sigma \ra             \, ,  \nn
\eqr
\noi where $\hat{\vsigma}$ denotes  denotes the vector of spin Pauli matrices and 
\bsub
\bqr
\rho_q (\br, \br')
& \equiv & \sum_{\sigma} \rho_q (\br \sigma ,\br' \sigma) \q ,  \\
\bs_q (\br,\br')
& \equiv & \sum_{\sigma \sigma'} \rho_q (\br \sigma ,\br' \sigma') \;
      \la \sigma' | \hat{\vsigma} | \sigma \ra       \q .      
\eqr
\esub
\noindent Below, we not only make use of quantities labelled by
$q=n,p$, but also of the associated isoscalar ($t=0$) and isovector ($t=1$) quantities. 
The former (latter) are obtained by taking the sum (difference) of corresponding 
neutron and proton quantities.

The standard Skyrme energy density functional (EDF) kernel derived from the interaction defined 
through Eq.~(\ref{skyrmestart}) can be expressed 
in terms of local densities and currents, and we refer the reader 
to Refs.~\cite{RMP,per04,Les07a} for more details. 
These are 
matter (scalar) density $\rho_q (\br)$, 
kinetic (scalar) density $\tau_q (\br)$, 
current (vector) density $\bj_q (\br)$, 
spin (pseudo-vector) density $\bs_q (\br)$, 
spin kinetic (pseudo-vector) density $\bT_q (\br)$, 
spin-current (pseudo-tensor) density  $J_{q,\mu \nu} (\br)$, 
and tensor-kinetic (pseudo-vector) density $\bF_{q} (\br)$ densities. 
Densities $\rho_q (\br)$, $\tau_q (\br)$ and $J_{q, \mu \nu}(\br)$
are even under time-reversal transformation while $\bs_q (\br)$, $\bT_q (\br)$,
$\bj_q (\br)$ and $\bF_{q} (\br)$ are odd and are defined as (to have a better notation the $q$ index is omitted in the following 
densities)
\bsub
\label{olddensities}
\bqr
\label{eq:locdensities:Jmunu}
J_{\mu \nu} (\br)
& = & \frac{1}{2i} \, \left( \nabla_{\mu} - \nabla_{\mu}^{\prime} \right) \; s_{\nu} (\br,\br')
      \big|_{\br = \br'}                                             \q ,  \\ 
\label{eq:locdensities:Tmu}
T_{\mu} (\br)
& = & \vnabla \cdot \vnabla^{\prime} s_{\mu} (\br,\br')
      \big|_{\br = \br'}                                             \q ,  \\ 
\bj_{\mu} (\br)
& = & -\frac{i}{2}(\nabla_\mu - \nabla_\mu^{\prime}) \; \rho (\br,\br')
      \big|_{\br = \br'}                                             \q ,  \\
\bF_{\mu} (\br)
& = & \frac{1}{2} (\nabla_\mu\nabla^{\prime}_\nu + \nabla_\mu^{\prime}\nabla_\nu) \; s_{\nu} (\br,\br')
      \big|_{\br = \br'}                                             \q . 
\eqr
\esub
Note that for the density $\bF_{\mu} (\br)$ written above, and for all formulas in this paper, the convention of an implicit sum for repeated indices is used. The $4th$ order contribution to the Skyrme pseudo-potential (Eq.~\ref{V-four})  requires the definition of new additional densities~\cite{Dob00a} 
\bsub
\label{newdensities}
\bqr
\label{eq:locdensities:taumunu}
\tau_{\mu \nu} (\br)
& = & \nabla_{\mu} \nabla^{\prime}_{\nu} \; \rho (\br,\br')
      \big|_{\br = \br'}                                             \q ,  \\
\label{eq:locdensities:Tmunukappa}
K_{\mu \nu \kappa}(\br)
& = & \nabla_{\mu} \nabla^{\prime}_{\nu} \; s_{\kappa} (\br,\br')
      \big|_{\br = \br'}                                             \q ,  \\
&   &                                                                  \nn \\       
\label{eq:locdensities:Pimu}
\Pi_{\mu} (\br)
& = & \vnabla \cdot \vnabla^{\prime} j_{\mu} (\br,\br')\nn
      \big|_{\br = \br'}                                                   \\
& = & \frac{1}{2i} \, \left( \nabla_{\mu} - \nabla_{\mu}^{\prime} \right) \; \tau (\br,\br')
      \big|_{\br = \br'}                                             \q ,  \\ 
V_{\mu \nu} (\br)
& = & \vnabla \cdot \vnabla^{\prime} J_{\mu \nu} (\br,\br')
      \big|_{\br = \br'}                             \nn                      \\
& = & \frac{1}{2i} \, \left( \nabla_{\mu} - \nabla_{\mu}^{\prime} \right) \; T_{\nu} (\br,\br')
      \big|_{\br = \br'}                                             \q ,  \\ 
&   &                                                                  \nn \\
\label{eq:locdensities:Q}
Q (\br)
& = & \Delta \, \Delta^{\prime} \; \rho (\br,\br')     
      \big|_{\br = \br'}                                             \q ,  \\ 
\label{eq:locdensities:S}
S_{\mu} (\br)
& = & \Delta \, \Delta^{\prime} \; s_{\mu} (\br,\br')   
      \big|_{\br = \br'}                                             \q .
\eqr
\esub
\noindent Similarly to the cartesian spin-current pseudo-tensor density $J_{q, \mu\nu}(\br)$, 
$\tau_{q, \mu \nu}(\br)$ can be decomposed into a pseudo-scalar, an anti-symmetric vector 
and a symmetric traceless pseudo-tensor part as
\begin{equation}
\label{decomponewdensities}
\tau_{q, \mu \nu} (\br)
=   \tfrac{1}{3} \delta_{\mu \nu} \, \tau^{(0)}_q (\br)
  + \tfrac{1}{2}\epsilon_{\mu\nu\kappa} \,
                 \tau^{(1)}_{q, \kappa} (\br)
  + \tau^{(2)}_{q, \mu\nu} (\br) \, ,
\end{equation}
\noi where $\delta_{\mu\nu}$ is the Kronecker symbol and $\epsilon_{\mu\nu\kappa}$
the Levi-Civita tensor.  
In terms of Cartesian components, one has
\bsub
\label{defdecompo}
\bqr
\label{eq:tau:recoupled}
\tau^{(0)}_q(\br)
& \equiv & \tau_{\mu\mu}(\br)             
                                                       \q ,   \\
\tau^{(1)}_{q, \kappa} (\br)
& \equiv & \epsilon_{\kappa \mu \nu} \, \tau_{q, \mu\nu} (\br)    
                                                       \q ,   \\
\tau^{(2)}_{q, \mu \nu} (\br)
& \equiv & \tfrac{1}{2} [ \tau_{q, \mu \nu}(\br) + \tau_{q, \nu \mu}(\br) ]
    - \tfrac{1}{3} \delta_{\mu\nu} \tau_{q, \kappa \kappa} (\br) 
                                                       \q .
\eqr
\esub
Contrary to $J_{q, \mu\nu}(\br)$, the vector part $\tau^{(1)}_{q, \kappa} (\br)$ 
is the only vanishing contribution when spherical symmetry is imposed. As we will see in the following section, the presence of rank-2 tensor in spherical symmetry will introduce major modifications to the mean field equations.

%

\subsubsection{The Skyrme energy density functional.}


We write the Skyrme part of the NEDF kernel in the more convenient form 
\bqr
\CE_{\text{Sk}} & \equiv & \CE^{\text{(0)}}_{\text{Sk}} + \CE^{\text{(2)}}_{\text{Sk}} 
                         + \CE^{(4)}_{\text{Sk}}      + \CE^{\text{LS}}_{\text{Sk}} + \CE^{\text{DD}}_{\text{Sk}}  \nn  \\
                & \equiv & \sum _{t=0,1}\int d\br \, \left[ \mathcal{H}^{(0)}_{t} (\br)  + \mathcal{H}^{(2)}_{t} (\br) 
                           + \mathcal{H}^{(4)}_{t} (\br) 
                         + \mathcal{H}^{\text{LS}}_{t} (\br)+ \mathcal{H}^{\text{DD}}_{t} (\br)
                                                     \right]                    \q , \nn                                                     
\eqr
\noi
where local energy densities have been introduced. 
Excluding $\CE^{\text{(4)}}_{\text{Sk}}$ one gets the \emph{standard} Skyrme functional, and the explicit expressions for the energy densities can be found in the literature~\cite{per04,Les07a}.  

The 4th order contribution to the functional kernel can be decomposed 
by analysing the behaviour of $\tau_{q, \mu \nu}$ and $T_{q, \mu \nu \kappa}$ under time-reversal~\cite{per04}. 
One can thus distinguish between the {\it even} part, that survives 
in time-reversal symmetric systems, and the {\it odd} part, that is non-zero only in time-reversal symmetry breaking systems. 
Omitting the dependence on the position vector, one has
\bsub
\bqr
\label{eq:fullEDF4}
\mathcal{H}^{(4)}_t & = & 
      \big[   \mathcal{H}_t^{\text{(4),even}}
            + \mathcal{H}_t^{\text{(4),odd}}
      \big]                                                  \q ,      \\
\label{eq:ef:DKe}
\mathcal{H}_t^{\text{(4),even}} 
& = & C^{(4) \Delta \rho}_t \left( \Delta \rho_t \right)^2
    + C^{(4) M \rho}_t \bbM_t^{\text{even}} [\rho_t]             
    + C^{(4) M s}_t \bbM_t^{\text{even}} [\bs_t]             \q ,   \\    
\label{eq:ef:DKo}
\mathcal{H}_t^{\text{(4),odd}} 
& = & C^{(4) \Delta s}_t \left( \Delta \bs_t \right)^2
    + C^{(4) M \rho}_t \bbM_t^{\text{odd}} [\rho_t]           
    + C^{(4) M s}_t \bbM_t^{\text{odd}} [\bs_t]              \q ,
\eqr
\esub
\noi where the relation between $4th$ order coupling constants and $4th$ order interaction parameters are given in~\citeAppendix{app:cpl}. 
The $t$-index is omitted in the  $\bbM$ expressions for the sake of simplicity. All the indices are summed over the $x,y,z$ coordinates.
\bsub
\bqr
\bbM^{\text{even}} [ \rho ] 
& = & \tfrac{1}{8} \, \left\{ \, \rho \, Q  \, + \, \tau^2 \, \right\} + \tfrac{1}{4} \,\left[ \tau_{\mu \nu} \tau_{\mu \nu} 
   \, - \, \tau_{\mu \nu} \nabla_{\mu} \nabla_{\nu} \rho \;  \right]                  \q ,  \\
\bbM^{\text{even}} [ \bs ] 
& = & - \tfrac{1}{8} \, \left\{ \, \left( \nabla_{\mu} J_{\mu \nu} \right)^2 
   \, + \, 4 J_{\mu \nu} V_{\mu \nu} \, \right\}                           \q ,     \\
\bbM^{\text{odd}} [ \rho ] 
& = & - \tfrac{1}{8} \, \left\{ \, \left( \vnabla \cdot \bj \right)^2  
   \, + \, 4 \, \bj \cdot \vPi \, \right\}                                 \q ,     \\
\bbM^{\text{odd}} [ \bs ] 
& = & \tfrac{1}{8} \, \left\{ \, \bs \cdot \bS  \, + \, \bT^2  \, \right\}
+  \tfrac{1}{4} \, \left( K_{\mu \nu \kappa} K_{\mu \nu \kappa}  
   \, - \, K_{\mu \nu \kappa} \nabla_{\mu} \nabla_{\nu} s_{\kappa} \right).
\eqr
\esub
Since in this section we focus on the ground state properties of even-even nuclei, we shall consider only the time even part $\mathcal{H}_t^{\text{D,even}}$ in the following. The time odd part $\mathcal{H}_t^{\text{D,odd}}$ will be explicitly taken into account in the section devoted to the properties of the Linear Response theory for this functional.

%
\subsubsection{The single-particle Hamiltonian.}
%
%

The isospin representation of the NEDF is convenient for a discussion of its physical content. Many of available Hartree-Fock or Hartree-Fock-Bogoliubov (HFB) codes use a proton-neutron representation~\cite{Bon87a} that is better suited to the construction of the one-body potentials and the symmetries chosen here. Thus, the total time-even part of the $4th$ order energy density is expressed as
\bwt
\bqr
\label{eq:fullEDF4_even_np}
\mathcal{H}^{(4),even} 
  =  & C^{(4) \Delta \rho}_- \, \left( \Delta \rho_q \right)^2  
 \, + \, 2 \, C^{(4) \Delta \rho}_1 \, \left( \Delta \rho_0 \right)^2                        \nn  \\
  & +   \tfrac{1}{8} C^{(4) M \rho}_-    
  \, \Big\{ \, \left[ \, \rho_0 \, Q_0  \, + \, \tau_0^2 \, \right]                 
 \, + \, 2 \, \left[ \tau_{0, \mu \nu} \tau_{0, \mu \nu} 
 \, - \, \tau_{0, \mu \nu} \nabla_{\mu} \nabla_{\nu} \rho_0 \right] \, \Big\}              \nn \\
  & +   \tfrac{1}{4} \, C^{(4) M \rho}_1 \, \Big\{ \, \left[ \, \rho_q \, Q_q \, + \, \tau_q^2 \, \right]                        
 \, + \,                2 \left[ \tau_{q, \mu \nu} \tau_{q, \mu \nu}        
 \, - \,  \tau_{q, \mu \nu} \nabla_{\mu} \nabla_{\nu} \rho_q \right] \, \Big\}             \nn \\
  & -   \tfrac{1}{8} \, C^{(4) M s}_-     
                      \, \left[ \, \left( \nabla_{\mu} J_{0, \mu \nu} \right)^2 
 \, + \, 4 J_{0, \mu \nu} V_{0, \mu \nu} \, \right]       \nn \\                                  
& -    \tfrac{1}{4} \, C^{(4) M s}_1 \, \left[ \, \left( \nabla_{\mu} J_{q, \mu \nu} \right)^2 
 \, + \, 4 J_{q, \mu \nu} V_{q, \mu \nu} \, \right] \, ,
\eqr
\ewt
\noindent where we have introduced a shorthand notation   $C_0^X-C_1^X \equiv C_-^X$, being $X=\Delta\rho, M\rho,\dots$
The equations of motion for proton and neutron single-particle states are obtained 
through standard functional derivative techniques~\cite{RMP,per04} and read
\be
\label{eq:mf:eq}
\hat{h}_q (\br) \, \psi_i(\vec{r}) = \varepsilon_i \, \psi_i (\br) \, .
\ee
The expression of the one-body hamiltonian $\hat{h}_q (\br)$ as obtained 
from the standard Skyrme functional has been given in Ref.~\cite{Bon87a}. 
In the present case, the $4th$ order contribution provides the effective mass 
with a tensor character (see~\citeqdot{eq:effmass:fullD}) such that 
the one-body Hamiltonian must be generalized to the form
\begin{equation*}
\fl \label{eq:sphamil}
\hat{h}_q (\br) = U_q (\br)
      + \Delta \Big( V_q (\br) \Delta \Big)
		  - \nabla_{\mu} \cdot B_{q, \mu \nu} (\br) \nabla_{\nu}  - \tfrac{i}{2}
        \big[ W_{q, \mu \nu} (\br) \, \nabla_\mu
              + \nabla_\mu \, W_{q, \mu \nu} (\br)
        \big] \, \hat{\sigma}_\nu\,,
\end{equation*}
with the following fields (we indicate explicitly  here only the $4th$ order contribution)
\bwt
\bqr
\label{eq:vqD}
V_q^{(4)} (\br) 
 & = &  \frac{\delta \CE^{(4)}_{\text{Sk}}}{\delta Q_q (\br)}          =   \tfrac{1}{8} \, C^{(4) M \rho}_- \, \rho_0  
\, + \, \tfrac{1}{4} \, C^{(4) M \rho}_1 \, \rho_q                     \,,           
\eqr
\bqr
\label{eq:uqD}
U_q^{(4)} (\br) 
 & = & \frac{\delta \CE^{(4)}_{\text{Sk}}}{\delta \rho_q (\br)}                \nn      \\            
 & = & 2 \, C^{(4) \Delta \rho}_-
        \, \Delta \Delta \rho_0
\, + \,    \tfrac{1}{4} \, C^{(4) M \rho}_-  
                    \, \Big\{ \, \tfrac{1}{2} Q_0 
                    - \nabla_\mu \nabla_\nu \tau_{0, \mu \nu}
                    \, \Big\}                   \nn \\
 & + & 4 \, C^{(4) \Delta \rho}_1 \, \Delta \Delta \rho_q                                          
\; + \; \tfrac{1}{2} \, C^{(4) M \rho}_1   
                     \, \Big\{ \, \tfrac{1}{2} Q_q 
                    - \nabla_\mu \nabla_\nu \tau_{q, \mu \nu}
                     \, \Big\}           \,,                                    
\eqr
and for $B_{q, \mu \nu}$
\bqr
\label{eq:effmass:fullD}
B_{q, \mu \nu} & \equiv & 
   \frac{\delta \CE_{\text{Sk}}}{\delta \tau_{\mu \nu} (\br)}    \nn   \\
& = & \Big\{ \, \frac{\hbar^2}{2m} 
        \; + \; C_-^{\tau} \rho_0 
        \; + \; 2 C_1^{\tau}  \rho_q \, + \, \tfrac{1}{4} \, C_-^{(4) M \rho} \, \tau_0 
        \; + \; \tfrac{1}{2} \, C_1^{(4) M \rho} \, \tau_q \, \Big\} \, \delta_{\mu \nu}    \nn \\
& + &           \tfrac{1}{4} \, C_-^{(4) M \rho} 
                             \, \left[ 2 \tau_{0, \mu \nu} - \nabla_{\mu} \nabla_{\nu} \rho_0 \right]  
\, + \,         \tfrac{1}{2} \, C_1^{(4) M \rho} 
                             \, \left[ 2 \tau_{q, \mu \nu} - \nabla_{\mu} \nabla_{\nu} \rho_q \right] \nn \\
& - & \tfrac{1}{2} \, C^{(4) M s}_{-} \, J_{0, \mu \nu}                
\, - \,                                       C^{(4) M s}_{1} \, J_{q, \mu \nu}\,.
\eqr
The spin-orbit field  $W_{q,\mu \nu} (\br)$ is also a tensor to which the $4th$ order contributes
\bqr
\label{eq:wqD} 
W_{q,\mu \nu}^{(4)} (\br) 
 & = & \frac{\delta \CE^{(4)}_{\text{Sk}}}{\delta J_{q,\mu \nu} (\br)}  \nn                  \\       
 & = & \tfrac{1}{4} \, C^{(4) M s}_{-} \,
                  \nabla_{\mu} \nabla_{\kappa} J_{0, \kappa \nu}                      
\, + \, \tfrac{1}{2} \, C^{(4) M s}_{1} \,
                  \nabla_{\mu} \nabla_{\kappa} J_{q, \kappa \nu}                      \nn \\
 & - & \tfrac{1}{2} \, C^{(4) M s}_{-}\, V_{0, \mu \nu}    
\, - \,                C^{(4) M s}_{1} \, V_{q, \mu \nu}.
\eqr
\ewt

\noindent We refer to Ref.~\cite{per04} for the expressions of the fields at second order.

%
\subsection{Spherical symmetry}
\label{sect:skyrmefu:sphere}
%
%

In this section, we enforce the spherical symmetry on the one-body Hamiltonian. This is of particular interest to perform calculations of semi-magic nuclei.
The single-particle wave functions solutions of Eq.~(\ref{eq:mf:eq}), 
from which the densities are built, are labelled by $(n, \ell, j, m, q)$, 
where $n$ denotes the principal quantum number, $\ell$ the orbital angular momentum, 
$j$ the total angular momentum, $m$ the projection of the latter on the $z$-axis, 
and $q$ the isospin projection. 
Wave functions separate into radial, angular and spin parts, the latter 
two making up a spherical harmonic tensor $\Omega_{\ell j m}({\hat r})$
\bqr
\psi_{n \ell j m q} (\br) 
            =  u_{n\ell j q} (r) \; \Omega_{\ell j m} ({\hat r}) = \frac{1}{r} R_{n\ell j q} (r) \; \Omega_{\ell j m} ({\hat r})  \q . 
\eqr
%
 After some tedious calculations, Eq.~(\ref{eq:mf:eq}) can be solved to determine the radial part of the wave function $R_{n\ell j q} (r)$ for each quantum number $\{n\ell j\}$ and it reads
\bsub
\begin{eqnarray}
A_4 R_{n\ell j}^{(4)} +  A_3 R_{n\ell j}^{(3)} + A_2  R_{n\ell j}^{(2)} + A_1 R_{n\ell j}^{(1)} + A_0 R_{n\ell j} = \epsilon_{n\ell j} R_{n\ell j} \, .
\end{eqnarray}
In the following equations a superindex $(i)$ on a radial function represents its $i$th derivative with respect to the radial coordinate $r$. The quantities $A_n$ are defined as 

\begin{eqnarray}
 A_4 &=&\tfrac{1}{8} \,  C^{(4) M \rho}_-  \, \rho_0  \, + \, \tfrac{1}{4} \, C^{(4) M \rho}_1 \, \rho_q \\ 
 A_3 &=&  \tfrac{1}{4} \,  C^{(4) M \rho}_-  \, \rho_0^{(1)}   +  \tfrac{1}{2} \, C^{(4) M \rho}_1 \, \rho_q^{(1)}\\
A_2  &=&  -  \frac{\hbar^2}{2m}  -   C_-^{\tau} \ \rho_0  +  2 C_1^{\tau} \rho_q  -   \tfrac{1}{4}   C_-^{(4) M \rho} \left[ 3 \tau_{0_R} +  \tau_{0_C}  - \frac{3}{2} \rho_0^{(2)} \right] \\
&-&   \tfrac{1}{2} C_1^{(4) M \rho}\left[ 3 \tau_{q_R} +  \tau_{q_C} - \frac{3}{2} \rho_q^{(2)} \right]   -  \frac{ \ell (\ell+1)}{r^2} \left[\tfrac{1}{8} \,  C^{(4) M \rho}_-  \, \rho_0  \, + \, \tfrac{1}{4} \, C^{(4) M \rho}_1 \, \rho_q \right] \nn \\
A_1 &=&  - C_-^{\tau} \rho_0^{(1)}  -  2 C_1^{\tau} \rho_0^{(1)}  \nn \\
&& + \tfrac{1}{4}   C_-^{(4) M \rho} \left[ 3 \tau_{0_R}^{(1)} +  \tau_{0_C}^{(1)} -  \rho_0^{(3)} \right] + \tfrac{1}{2} C_1^{(4) M \rho}\left[ 3 \tau_{q_R}^{(1)} +  \tau_{q_C}^{(1)} -  \rho_q^{(3)} \right]  \nn \\ 
&& +  \frac{ \ell (\ell+1)}{r^2} \left[\tfrac{1}{4} \,  C^{(4) M \rho}_-  \, \left(\rho_0^{(1)}-2\frac{\rho_0}{r}\right)  \, + \, \tfrac{1}{2} \, C^{(4) M \rho}_1 \, \left(\rho_q^{(1)} - 2\frac{\rho_q}{r}\right) \right] 
\end{eqnarray}
\begin{eqnarray}
A_0= &&  U_q (r)+ W_q (r)  \nn \\
           &+&   C_-^\tau \frac{\rho_0^{(1)}}{r} + 2 C_1^\tau \frac{\rho_q^{(1)}}{r} + \frac{\ell (\ell+1)}{r^2} \left[ - \frac{\hbar^2}{2m} + C_-^\tau \rho_0 + 2 C_1^\tau \rho_q \right] \nn \\
           &+& \tfrac{1}{4} \,  C^{(4) M \rho}_- \left[ 3 \frac{ \tau_{0_R}}{r} - \frac{\rho_0^{(3)}}{r} \right] + \tfrac{1}{2} \,  C^{(4) M \rho}_1 \left[ 3 \frac{ \tau_{q_R}}{r} - \frac{\rho_q^{(3)}}{r} \right] \nn \\
           &-& \frac{\ell (\ell+1)}{4r^2} C^{(4) M \rho}_- \left[ \tau_{0_R}  + \frac{1}{2} \rho_0^{(2)}  + 3 \frac{\rho_0^{(1)}}{r} + 3 \frac{\rho_0}{r^2} \right] \nn \\
            &-& \frac{\ell (\ell+1)}{2r^2} C^{(4) M \rho}_1 \left[ \tau_{q_R} + \frac{1}{2} \rho_q^{(2)} + 3 \frac{\rho_q^{(1)}}{r}+ 3 \frac{\rho_q}{r^2} \right] \nn \\
           & + &  \frac{\ell^2 (\ell+1)^2}{8 r^4} C^{(4) M \rho}_-  \rho_0 + \frac{l^2 (l+1)^2}{4r^4} C^{(4) M \rho}_1  \rho_q\,.
\end{eqnarray}
\esub
We recall that the two scalar fields used in previous expressions read
\bqr
\label{eq:uqD}
U_q (r) 
 & = & \frac{\delta \CE_{\text{Sk}}}{\delta \rho_q (r)}   \\            
 & = & U^{Sk}_q(r) + 2 \, C^{(4) \Delta \rho}_-   \Delta \Delta \rho_0 +  4 \, C^{(4) \Delta \rho}_1 \, \Delta \Delta \rho_q                       \nn  \\
 & + & \tfrac{1}{4} \, C^{(4) M \rho}_- 
                    \, \Big\{ \, \tfrac{1}{2} Q_0 
                    - \nabla_\mu \nabla_\nu \tau_{0, \mu \nu}
                    \, \Big\} + \tfrac{1}{2} \, C^{(4) M \rho}_1   
                     \, \Big\{ \, \tfrac{1}{2} Q_q 
                    - \nabla_\mu \nabla_\nu \tau_{q, \mu \nu}
                     \, \Big\} \nn
\eqr
%
%
%
\begin{equation}
\fl W_q (r) = W^{Sk}_q(r) \; + \; \left( j (j+1) - \ell (\ell+1) -\frac{3}{4} \right)  \left[ \frac{3}{4} C^{(4) M s}_- \frac{V_0(r)}{r} + \frac{3}{2}  C^{(4) M s}_{1} \frac{V_q(r)}{r} \right],
\end{equation}

\noindent where $U^{Sk}_q(r),W^{Sk}_q(r)$ are the central and spin-orbit fields for the standard Skyrme functional up to second order~\cite{Cha97a}.
The local density are now expressed in spherical symmetry as
\begin{equation}
\rho_{q,0} (r)                = \salpha \! \wfusqalpha
\end{equation}
where $\{\alpha\} \equiv \{nlj\}$ ($\{nljq\}$) if the index of the density is $q$ ($0$). The summation is limited over $\{\alpha\}$ states below the Fermi energy, and $u_{\{\alpha\}} \equiv R_{\{\alpha\}}/r$.  Furthermore, one has ($X_\mu$ represent the usual cartesian coordinates)
\begin{eqnarray}
\label{eq:tauRC}
\label{eq:tau}
\fl & \tau_{q,0,\mu\nu}   (r)  =  \frac{1}{2}\tau_{q,0,C} \; \delta_{\mu\nu} \, + \, \frac{X_\mu X_\nu}{r^2}\left[\tau_{q,0,R} +\frac{1}{2}\tau_{q,0,C}\right]        \q ,     \\
\fl \label{eq:tauR}
\fl & \tau_{q,0,R} (r) =  \salpha \wfupsqalpha  \q ,     \\
\label{eq:tauC}
\fl & \tau_{q,0,C} (r) =  \salpha
               \ell ( \ell + 1) \; \frac{\wfusqalpha}{r^2}     \q ,   \\
\nn
\fl & V_{q,0}(r)= \salpha  \left[ \frac{\wfusqalpha}{r^3} \ \left[ 1 -  \ell(\ell+1)  \right] - \frac{\wfupsqalpha }{r} \right] \; \left[ j(j+1) - \ell(\ell+1) - \frac{3}{4} \right] \\
\fl & J_{q,0}(r) =  \salpha   \left[ j(j+1) - \ell(\ell+1) - \frac{3}{4} \right] \frac{\wfusqalpha}{r} \q , \\
\fl & Q_{q,0}(r) =  \salpha  \left[\Delta u_{\{\alpha\}}(r)- \ell(\ell+1)\frac{u_{\{\alpha\}}(r)}{r^2}\right]^2 \q .
\end{eqnarray}
Contrary to the standard Skyrme potential, the differential equation is now of fourth order, but no particular other difficulty appears.


\section{Linear response for $4th$ order component}\label{LR:skyrme}


In a recent series of articles, we have presented the Linear Response formalism~\cite{gar92,Dav09a,pas12,pas12b} for a standard Skyrme functional in both symmetric nuclear matter (SNM) and pure neutron matter (PNM). In particular, using the LR formalism we have studied the presence of finite-size instabilities in the infinite medium. The presence of these modes can be related to the presence of analogous instabilities in finite nuclei~\cite{Les06a,pas12c,hel12,sch10}. 
In ref.~\cite{hel13}, we have performed a systematic study, although limited to the scalar-isovector channel, of these instabilities  showing that they arise from a badly constrained the coupling constant that multiplies gradient terms. In the same article, we have also derived a quantitative criterion to detect these instabilities using the simple LR formalism in the infinite medium.
Due to its very low computational  cost, the LR  formalism can be directly included into the optimization procedure used to determine the coupling constants of the functional so to avoid the exploration of regions of parameters that can not produce stable functionals.
In Ref.~\cite{pas13c}, we have presented for the first time a new fitting procedure based on the LR formalism to produce stable Skyrme functionals.

In the present section, we  extend  the LR formalism for a $standard$ Skyrme functional in SNM to include $4th$ order terms. Since we want to focus here mainly on the role of these higher  order terms, we will neglect the explicit tensor contribution.
Before discussing the details of the response function, we have to briefly mention the modifications induced by these extra  terms into the effective mass, which is defined as~\cite{fet71}
\begin{equation}
\frac{1}{m^* (k)}= \frac{1}{k}\frac{dU(k)}{dk}\, ,
\end{equation}
where $U(k)$ is the mean-field potential and $k$ is the impulsion of the particle. Using the expression of the complete Skyrme functional including higher order terms, we have
\be\label{eff:mass}
   \left( \frac{m}{m^*} \right)_{(0,0)} \, = \, 1 \, + \, \frac{2m}{\hbar^2} \, \rho_0 
                                        \, \left[ C_0^\tau + \tfrac{1}{4} (k_F^2 +k^2)C_0^{(4) M \rho}
                                        \, \right]                                       \q .
\ee
For Skyrme's original pseudo-potential, $i.e.$ up to $2nd$ order, there is no explicit momentum dependence. In fact, the highest order contribution is in $k^2$: the derivative together with the factor $1/k$ eliminates all momentum dependence. When $4th$ order is added, we find terms in $k^4$ and things are thus differents. 

 We already mentioned that this $4th$ order pseudo-potential has actually to be considered as a polynomial expansion in terms of gradients of a finite-range potential. It is thus not surprising to recover one fundamental aspect of any finite range pseudo-potential, that is the momentum dependence of the effective mass.
  In nuclear physics, since all energy scales are below the Fermi energy, it is traditional to take $k=k_F$ in the above equation~\cite{heb09}. 
  Eq.(\ref{eff:mass}), can be re-expressed in terms of pseudo-potential coefficients as
\be
\fl   \left( \frac{m}{m^*} \right)_{(0,0)} \, = \, 1 \, + \, \frac{2m}{\hbar^2} \, \rho_0 
                                        \, \left[ \frac{1}{16} (3 t_1 + t_2 (5 + 4 x_2))
                                        + \tfrac{1}{16} (k_F^2 +k^2)(3 t_1^{(4)} + t_2^{(4)} (5 + 4 x_2^{(4)})) \, \right]     \,.      
\ee

  Qualitatively, it has been noticed in Ref.~\cite{dav14b}, that the interaction parameters is one order of magnitude smaller between two orders (see also discussion in Ref.~\cite{kor10R}). For simplicity we will therefore consider $t_1^{(4)} \simeq t_1/(10 k_F^2)$ and the same for the other parameters. Thus, replacing $k$ by $k_F$ is a good approximation only when
\be
(k_F^2 +k^2)/(10 k_F^2) \ll 1\,,
\ee
that is $k\simeq 3k_F$. In the following we will present explicitly an illustration of the effect of our approximation through the energy-weighted sum rule (EWSR).

The advantage of this approximation is that we can strongly simplify the expressions of particle-hole (ph) propagators. With this proviso in mind, we can then generalize our formalism in a straightforward way.
All the ingredients and formulas are given in~\ref{beta functions}, in particular the generalized Linhardht functions as well as the $\beta_{i}(q,\omega)$ functions (notations and conventions are those of Ref.~\cite{Dav09a,pas12,pas12b,pas14}) entering in the resolution of Bethe-Salpeter equations are given explicitly.

Solving the the Bethe-Salpeter~\cite{gar92}  equations in SNM, we obtain the response function $\chi^{(\alpha)}(q,\omega)$ of the system in each channel $\alpha$, where $\alpha=(S,M,I)$ is a shorthand notation for the quantum number of the system:  $S(I)$ is the total spin (isospin) and $M$ is the spin projection along the $z$-axis. 
 
Since the number of coupled equations has largely increased as compared to the case shown in Refs.\cite{Dav09a,pas12,pas12b,gar92}, we have decided to express the system of coupled equations in matrix form as done in Ref.~\cite{pas14} and to solve them numerically to obtain the response function of the system $\chi^{(\alpha)}(q,\omega)$. 
The instabilities in SNM can thus be found as the numerical solutions of
\begin{eqnarray}
1/\chi^{(\alpha)}(\omega=0,q)=0\,.
\end{eqnarray}

From the matrix form, it is also possible to take explicitly the limit $\omega \rightarrow \infty$ and get the energy-weighted sum rule $M_1$.
The explicit expression reads
\begin{eqnarray}\label{m1:analytic}
M_{1}^{(S,M,I)}/N=\frac{q^{2}}{2m^{*}}\left[ 1-\frac{m^{*}\rho}{2}\left(W_{2}^{(S)}+\left[4k_{F}^{2}+q^{2}\right]W_{4}^{(S)} \right)\right]\,.
\end{eqnarray}
%

\begin{figure}[!h]
\begin{center}
        \includegraphics[clip,scale=0.34,angle=-90]{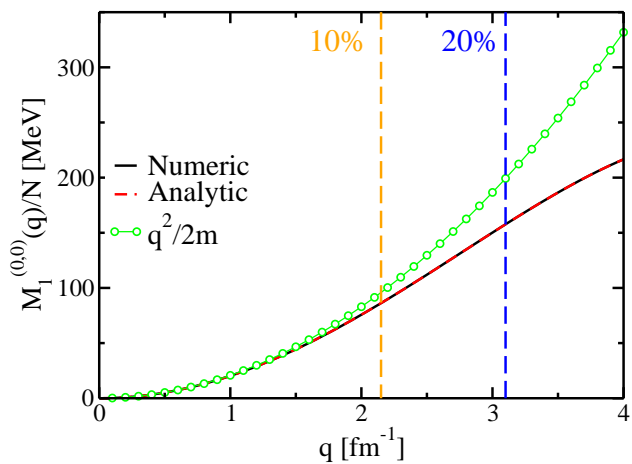}
\caption{(Color online) The EWSR in the (0,0) channel at $\rho=0.16$ fm$^{-3}$ for the modified version of SLy5, see text for details. The solid and dashed line correspond to the numerical and analytical calculations of the $M_{1}^{(0,0)}$ using the LR formalism. The symbols represent the result obtained with double commutator techniques. The vertical dashed lines represent the deviation between the two methods. }
\label{fig:m1}
\end{center}
\end{figure}

In Fig.\ref{fig:m1}, we show the EWSR at $\rho=0.16$ fm$^{-3}$ obtained in our LR code either using the numerical integration or the analytic expansion (see Ref.~\cite{gar92} for more details) in the (0,0) channel. Since there are no available parameterizations of the N2LO functional obtained from a consistent fitting procedure, we have taken the SLy5 Skyrme functional~\cite{Cha97a,Cha98a} and for the fourth order term we have taken $1/10$ of the second order value (no change in the $x_{i=1,2}^{(4)}$) parameters.

We notice that the analytic expansion and the numerical integration stay on top of each other as expected, demonstrating the validity of our calculations.
The EWSR  can also be calculated using double commutator techniques~\cite{boh79} as already explained in Ref.~\cite{pas12}. The calculation with this method in the  channel (0,0) is particularly simple. In this case, the excitation operator is actually the plane wave $e^{i\mathbf{q}\mathbf{r}}$. From the physical point of view such operator represent a translation. Since our pseudo-potential is by construction Galilean invariant~\cite{dav13}, it commutes with this operator. As a result we are left with the kinetic operator only and the EWSR reads $M_{1}^{(0,0)}=q^{2}/2m$. The result is shown in Fig.\ref{fig:m1}.
We notice that the approximation on the momentum dependence of the effective mass $m^{*}(k=k_{F})$ leads to a discrepancy less of 10\% up to $q\approx2$ fm$^{-1}$, while at around $q\approx$3 fm$^{-1}$ the relative error grows up to 20\%.
Since the EWSR can be calculated analytically we can show that the discrepancy arises from the term

\begin{eqnarray}
M_{1}^{(0,0)}/N-\frac{q^{2}}{2m}=-\frac{1}{8}C^{(4)M\rho}_{0}q^{4}\rho\,,
\end{eqnarray}

\noindent where $M_{1}^{(0,0)}$ has been defined in Eq.\ref{m1:analytic}. We immediately observe that the differences between the two approaches grows as $q^{4}$.



\section{Summary and conclusions}
\label{conclusions}

We have discussed the contribution of $4th$ order terms to $standard$ Skyrme pseudo-potential.
In particular, we have studied how the corresponding single-particle equations are modified by the presence of these higher derivative terms. The functional formalism have been worked out at first in Cartesian basis and then specialized for the case of spherical symmetry in view of a future fit.
We have also presented the extension of the formalism of the Linear Response theory in symmetric nuclear matter to take into account these extra terms. The LR formalism has been shown~\cite{hel13} to be very useful to detect finite-size instabilities and it can be also included directly into the optimization procedure used to determine coupling constants~\cite{pas13c}.
With the tools presented in the present article together with some important ground state properties of the infinite medium already discussed in Refs.~\cite{dav13,dav14b}, it is now possible to fit the coupling constants of the functionals.



\section*{Acknowledgments}


We thank J.~Dobaczewski for stimulating discussions, T. Duguet, M. Bender and K. Bennaceur for discussions in the earliest stage of this work and Isaac Vida\~na for clarifying discussions on the role of the effective mass in Brueckner-Hartree-Fock calculations. The work of J.N. has been supported by grant FIS2011-28617-C02-2, Mineco (Spain). 


\begin{appendix}


\section{Coupling constants of the fourth-order part of the Skyrme EDF}
\label{app:cpl}

The $4th$ order coupling constants of the Skyrme functional can be expressed in terms of the 
parameters of the corresponding pseudo-potential by

\begin{table}[H]
\bc 
\label{tab:Skyrme_int:2bodyEDF:fourthorder_C:coeff}
\caption{Coefficients of the normal part of the fourth order functional, Eq.~(\ref{V-four}), 
 as a function of the parameters of the pseudo-potential of Eq.~(\ref{eq:fullEDF4_even_np}). 
Missing entries are zero.}  

\begin{tabular}{r cc cc cc cc }
                           &&                  &&                 &&                  &&                 \\[0.3mm] 
\hline \hline \noalign{\smallskip}
 && $t_1^{(4)}$ && $t_1^{(4)} x_1^{(4)}$ && $t_2^{(4)}$ && $t_2^{(4)} x_2^{(4)}$                         \\
\noalign{\smallskip}  \hline \noalign{\smallskip}
$C_0^{(4)\Delta\rho} \; =$ && $+\frac{9}{128}$ &&                 && $-\frac{5}{128}$ && $ -\frac{1}{32}$ \\[0.3mm]
$C_1^{(4)\Delta\rho} \; =$ && $-\frac{3}{128}$ && $-\frac{3}{64}$ && $-\frac{1}{128}$ && $-\frac{1}{64}$ \\[0.3mm]
$C_0^{(4) M \rho} \; =$    && $+\frac{3}{4}$   &&                 && $+\frac{5}{4}$   && $1$             \\[0.3mm]
$C_1^{(4) M \rho} \; =$    && $-\frac{1}{4}$   && $-\frac{1}{2}$  && $+\frac{1}{4}$   && $+\frac{1}{2}$  \\[0.3mm]
                           &&                  &&                 &&                  &&                 \\[0.3mm] 
$C_0^{(4)\Delta s} \; =$   && $-\frac{3}{128}$ && $+\frac{3}{64}$ && $-\frac{1}{128}$ && $-\frac{1}{64}$ \\[0.3mm]
$C_1^{(4)\Delta s} \; =$   && $-\frac{3}{128}$ &&                 && $-\frac{1}{128}$ &&                 \\[0.3mm]
$C_0^{(4) M s} \; =$       && $-\frac{1}{4}$   && $ \frac{1}{2}$  && $+\frac{1}{4}$   && $+\frac{1}{2}$  \\[0.3mm]
$C_1^{(4) M s} \; =$       && $-\frac{1}{4}$   &&                 && $+\frac{1}{4}$   &&                 \\[0.3mm]
\noalign{\smallskip} \hline \hline
\end{tabular}
\ec
\end{table}


\section{Beta functions}
\label{beta functions}

The 4th order pseudo-potential requires extra $\beta_i(q,\omega)$ functions for the calculation of the response function of the infinite medium. In the following we give the expressions of these new functions. The notations are those of Ref.\cite{gar92}.
\be
\label{app:beta:betafunct}
\beta_{i=9,14}(q,\omega) =\int \frac{d^{3}k}{(2\pi)^{3}}G_{HF}(\mathbf{k},\mathbf{q},\omega)F_{i=9,14}(\mathbf{k},\mathbf{q})
\ee
with
\begin{eqnarray}
F_{i=9,14}(\mathbf{k},\mathbf{q})&\equiv&\frac{k^{6}}{q^{6}},\frac{k^{8}}{q^{8}}, \frac{k^{4}(\mathbf{k}\cdot\mathbf{q})}{q^{6}},\frac{k^{4}(\mathbf{k}\cdot\mathbf{q})^{2}}{q^{8}},\frac{k^{2}(\mathbf{k}\cdot\mathbf{q})^{3}}{q^{8}},\frac{k^{6}(\mathbf{k}\cdot\mathbf{q})}{q^{8}}.
\end{eqnarray}
To do this calculations we have to introduce  higher generalized Lindhardt functions
\begin{eqnarray}
\fl &\Pi _{6} = -\frac{m  k_{F}^{2}}{16 \pi^{2}}\left[ 7+k ^{6}-\nu ^{2}-\nu ^{4}-\nu ^{6}+\frac{257}{3}k ^{4}+29\nu ^{2}k ^{4}+\frac{167}{3}k ^{2}+\frac{290}{3}k^{2}\nu ^{2}-29\nu ^{4}k ^{2}\right. \nonumber\\
&\; \; \; \; \; \; \; \; \; \; \; \; \; \; \; \; +(1+(k -\nu )^{2}) (1+k ^{4}-4k\nu +30k ^{2}\nu ^{2}+\nu ^{4}) A_{+}(k ,\nu )\nonumber\\
&\; \; \; \; \; \; \; \; \; \; \; \; \; \; \; \; +  \left.(1+(k +\nu )^{2}) (1+k ^{4}+4k\nu +30k ^{2}\nu ^{2}+\nu ^{4}) A_{-}(k ,\nu )\right] ,\\ \nonumber
\fl &\Pi _{8}= -\frac{m  k_{F}^{2}}{20 \pi^{2}}\left\{9+\frac{389}{3}k ^{2}+\frac{2561}{5}k ^{4}+427k ^{6}+k ^{8}-\nu ^{2}+268k ^{2}\nu ^{2}+\frac{1331}{3}k ^{4}\nu ^{2} \right.\nonumber\\
& + 46k ^{6}\nu ^{2} - \nu ^{4}-\frac{145}{3}k ^{2}\nu ^{4}+256k ^{4}\nu ^{4}-\nu ^{6}-46k ^{2}\nu ^{6}-\nu ^{8} \nonumber\\
&+\left[ k ^{8}-2k ^{7}\nu  +48 k ^{6}\nu ^{2}+k ^{6} -94 k ^{5}\nu ^{3} -4 k ^{5}\nu  +350 k ^{4}\nu ^{4}\right. \nonumber\\
&+\left. 55k ^{4}\nu ^{2}+k ^{4}-94 k ^{3}\nu ^{5}-200k ^{3}\nu ^{3}-6k ^{3}\nu  +48k ^{2}\nu ^{6}+55k ^{2}\nu ^{4}+66k ^{2}\nu ^{2}+k ^{2}-2k\nu ^{7}\right.\nonumber\\
&-\left.4k \nu ^{5}-6k \nu ^{3}-8k \nu +\nu ^{8}+\nu ^{6}+\nu ^{4}+\nu ^{2}+1\right]A_{+}(k ,\nu ) \nonumber\\
&+ \left[ k ^{8}+2k ^{7}\nu  +48 k ^{6}\nu ^{2}+k ^{6} +94 k^{5}\nu ^{3} +4 k ^{5}\nu  +350 k ^{4}\nu ^{4}\right. \nonumber\\
&+\left. 55k ^{4}\nu ^{2}+k ^{4}+94 k^{3}\nu ^{5}+200k ^{3}\nu ^{3}+6k ^{3}\nu  +48k^{2}\nu ^{6}+55k ^{2}\nu ^{4}+66k ^{2}\nu ^{2}+k ^{2}+2k\nu ^{7}\right.\nonumber\\
&+\left. \left.4k \nu ^{5}+6k \nu ^{3}+8k\nu +\nu ^{8}+\nu ^{6}+\nu ^{4}+\nu ^{2}+1\right] A_{-}(k ,\nu ) \right\}\,,
\end{eqnarray}
from which we can deduce
\begin{eqnarray}
\fl &\beta_{9}
=\frac{1}{64k ^{6}}\left[ \Pi _{6}-6 k \nu  \Pi _{4} +16 k ^{3}\nu ^{3}\Pi _{0}-\frac{8k ^{3}m^{*}  k _{F}\nu }{3\pi^{2}}\right] \nonumber \\ 
\fl & \beta_{10}
=\frac{1}{256k ^{8}} \left[ \Pi _{8}-8k \nu \Pi_{6}+64 k ^{3}\nu ^{3}\Pi _{2}-\frac{32 k ^{3} k _{F}m^{*} \nu }{3\pi^{2}}(1+2k ^{2}) \right]  \nonumber\\
\fl & \beta_{11}
=\frac{1}{32k ^{5}}\left[4k\nu (k -\nu )\Pi _{2}+(\nu -k )\Pi _{4}+\frac{2k  (1+2k ^{2})m^{*}  k _{F}}{3\pi^{2}}  \right]  \nonumber \\
\fl & \beta_{12}
=\frac{1}{64k ^{6}}\left[ (\nu -k )^{2}\Pi _{4}-4k \nu  (\nu -k )^{2}\Pi _{2} +\frac{k _{F}m^{*} (-5-76k ^{2}-120k ^{4}+20k \nu +40k ^{3}\nu )}{30\pi^{2}}\right]  \nonumber \\
\fl & \beta_{13}
=\frac{1}{32k ^{5}} \left[2k \nu  (k -\nu )^{3}\Pi _{0} - (k -\nu )^{3}\Pi _{2}+\frac{k _{F}m^{*} }{30\pi^{2}}(-5\nu +21k +70k ^{3}-40k ^{2}\nu +10k \nu ^{2}) \right]  \nonumber \\
\fl &\beta_{14}=\frac{1}{128k ^{7}}\left[ (\nu -k )\Pi _{6}+6k \nu (k -\nu )\Pi _{4}+16k ^{3}\nu ^{3}(\nu -k )\Pi _{0}\right.\nonumber \\  & \; \; \; \; \; \; \; \; \; \; \; \; \; \; \; \; \; \; \; \; \; \; \; \; \;  \left. +\frac{k  k _{F}m^{*} }{15\pi^{2}}\left(15+84k ^{2}+80k ^{4}+40k ^{3}\nu -40k ^{2}\nu ^{2} \right) \right]. 
\end{eqnarray}


\section{System of equations in each spin-isospin channel}
\label{systems}

\noindent Since the residual interaction is diagonal in isospin, all the Bethe-Salpeter equations are decoupled with respect to this quantum number. Notice that in the $S=0$ channel we have just only one spin projection $M$, thus we can neglect this index in this case in favor of a lighter notation.
\subsection{Channel $(S,I)=(0,I)$}

The unknown quantities entering in the system of equations are
%

\begin{tabular}{l l}
$X_{0}= \langle G_{RPA}^{(0,I)} \rangle$ & $X_{1} = \langle k^{2}G_{RPA}^{(0,I)} \rangle$\\
& \\
$X_{2}=\sqrt{\frac{4\pi}{3}}\langle k Y_{10}G_{RPA}^{(0,I)} \rangle$ & $X_{3}=\langle k^{4} G_{RPA}^{(0,I)} \rangle$\\
& \\
$X_{4}=\sqrt{\frac{4\pi}{3}} \langle k^{3} Y_{10}G_{RPA}^{(0,I)} \rangle$ & $X_{5}=\frac{4\pi}{3}\langle k^{2} Y_{10}^{2}G_{RPA}^{(0,I)} \rangle$\\
& \\
$Y_{1}=\sqrt{\frac{4\pi}{3}}\sum_{M'}M'\langle k Y_{1M'}G_{RPA}^{(1,M',I)}\rangle$ & $Y_{2}=\sqrt{\frac{4\pi}{3}}\sum_{M'} M' \langle k^{3} Y_{1M'}G^{(1,M',I)}_{RPA}\rangle$\\
& \\
$Y_{3}=\frac{4\pi}{3}\sum_{M'} M' \langle k^{2} Y_{1M'}Y_{10}G^{(1,M',I)}_{RPA}\rangle$ & 
\end{tabular}

\hfill 

\noindent which form together the vector denoted as $X_{(0,I)}$ in the following. From the Bethe-Salpeter equation, one can obtained after some straightforward calculations a system written in a matrix form as $A_{(0,I)} X_{(0,I)} =B_{(0,I)}$. For the sake of clarity, we decompose the matrix $A_{(0,I)}$ as 2 columns matrix of size $3\times9$ and $6\times9$ respectively $A_{(0,I)}=\left( A_{1} \; , \; A_{2} \right)+I_{9}$, where $I_{9}$ is the $9 \times 9$ identity matrix. The matrices $A_{1}$ and $A_{2}$ read

{\scriptsize
\begin{eqnarray}
\fl \left( \begin{array}{c c c c c }
-\beta_{0}W_{1}^{(0,I)}-q^{2}\beta_{2}W_{2}^{(0,I)}-q^{4}\beta_{5}W_{4}^{(0,I)} & &-\beta_{0}W_{2}^{(0,I)}-2W_{4}^{(0,I)}q^{2}(2\beta_{2}-\beta_{3})&& 2W_{2}^{(0,I)}q\beta_{1}+4 W_{4}^{(0,I)}q^{3}\beta_{4} \\
& & & &  \\
 -q^{2}\beta_{2}W_{1}^{(0,I)} -q^{4}\beta_{5} W_{2}^{(0,I)}-q^{6}\beta_{9}W_{4}^{(0,I)}& & -q^{2}\beta_{2}W_{2}^{(0,I)} -2W_{4}^{(0,I)}q^{4}(2\beta_{5}-\beta_{8}) & &  2W_{2}^{(0,I)}q^{3}\beta_{4}+4W_{4}^{(0,I)}q^{5}\beta_{11} \\
 & & & &  \\
-q\beta_{1}W_{1}^{(0,I)}-q^{3}\beta_{4}W_{2}^{(0,I)}-q^{5}\beta_{11}W_{4}^{(0,I)}  & & -q\beta_{1}W_{2}^{(0,I)}  -2W_{4}^{(0,I)}q^{3}(2\beta_{4}-\beta_{6})&&2W_{2}^{(0,I)}q^{2}\beta_{3}+4W_{4}^{(0,I)}q^{4}\beta_{8}\\
 & & & &  \\
-q^{4}\beta_{5}W_{1}^{(0,I)} -q^{6}\beta_{9}W_{2}^{(0,I)}-q^{8}\beta_{10}W_{4}^{(0,I)} & & -q^{4}\beta_{5}W_{2}^{(0,I)} -2W_{4}^{(0,I)}q^{6}(2\beta_{9}-\beta_{12}) && 2W_{2}^{(0,I)}q^{5}\beta_{11} +4W_{4}^{(0,I)}q^{7}\beta_{14}\\
 & & & &  \\
 -q^{3}\beta_{4}W_{1}^{(0,I)} -q^{5}\beta_{11}W_{2}^{(0,I)}-q^{7}\beta_{14}W_{4}^{(0,I)} & &  -q^{3}\beta_{4}W_{2}^{(0,I)} -2W_{4}^{(0,I)}q^{5}(2\beta_{11}-\beta_{13}) & &2W_{2}^{(0,I)}q^{4}\beta_{8} +4W_{4}^{(0,I)}q^{6}\beta_{12}\\
 & & & &  \\
-q^{2}\beta_{3}W_{1}^{(0,I)} -q^{4}\beta_{8} W_{2}^{(0,I)}-q^{6}\beta_{12}W_{4}^{(0,I)} && -q^{2}\beta_{3}W_{2}^{(0,I)} -2W_{4}^{(0,I)}q^{4}(2\beta_{8}-\beta_{7}) &&2W_{2}^{(0,I)}q^{3}\beta_{6} +4W_{4}^{(0,I)}q^{5}\beta_{13}\\
 & & & &  \\
4q^{3}(\beta_{2}-\beta_{3})C^{\nabla J}_{I} && 0 && 0\\
 & & & &  \\
4q^{5}(\beta_{5}-\beta_{8})C^{\nabla J}_{I} && 0 && 0\\
 & & & &  \\
4q^{4}(\beta_{4}-\beta_{6})C^{\nabla J}_{I}&& 0 && 0\end{array}\right)\nonumber\\
\end{eqnarray}
}

{\scriptsize
\begin{eqnarray}
\fl \left( \begin{array}{c  c  c  c  c  c}
 -\beta_{0}W_{4}^{(0,I)}&4W_{4}^{(0,I)}q\beta_{1} &2W_{4}^{(0,I)}q^{2}(\beta_{2}-3\beta_{3})&4q\beta_{0}C^{\nabla J}_{I} & 0 & 0\\
 & &  & &  &\\
 -q^{2}\beta_{2}W_{4}^{(0,I)} & 4W_{4}^{(0,I)}q^{3}\beta_{4}&2W_{4}^{(0,I)}q^{4}(\beta_{5}-3\beta_{8})&4q^{3}\beta_{2}C^{\nabla J}_{I} & 0 & 0\\
 & &  & &  &\\
-q\beta_{1}W_{4}^{(0,I)}&4W_{4}^{(0,I)}q^{2}\beta_{3}&2W_{4}^{(0,I)}q^{3}(\beta_{4}-3\beta_{6}) &4q^{2}\beta_{1}C^{\nabla J}_{I} & 0 & 0 \\
 & &  & &  &\\
-q^{4}\beta_{5}W_{4}^{(0,I)}&4W_{4}^{(0,I)}q^{5}\beta_{11}&2W_{4}^{(0,I)}q^{6}(\beta_{9}-3\beta_{12})&4q^{5}\beta_{5}C^{\nabla J}_{I} & 0 & 0 \\
  & &  & &  &\\
-q^{3}\beta_{4}W_{4}^{(0,I)} & 4W_{4}^{(0,I)}q^{4}\beta_{8}&2W_{4}^{(0,I)}q^{5}(\beta_{11}-3\beta_{13}) & 4q^{4}\beta_{4}C^{\nabla J}_{I} & 0 & 0\\
 & &  & &  &\\
-q^{2}\beta_{3}W_{4}^{(0,I)} & 4W_{4}^{(0,I)}q^{3}\beta_{6}&2W_{4}^{(0,I)}q^{4}(\beta_{8}-3\beta_{7}) & 4q^{3}\beta_{3}C^{\nabla J}_{I} & 0 & 0\\
 & &  & &  &\\
 0& 0 &0 & W_{2}^{(1,I)}q^{2}(\beta_{2}-\beta_{3}) & 2W_{4}^{(1,I)}q^{2}(\beta_{2}-\beta_{3}) & 4W_{4}^{(1,I)}q^{3}(\beta_{6}-\beta_{4})\\
  & &  &+2W_{4}^{(1,I)}q^{4}(\beta_{5}-\beta_{8}) &  &\\
  & &  & &  &\\
0& 0&0 &+W_{2}^{(1,I)}q^{4}(\beta_{5}-\beta_{8})  &2W_{4}^{(1,I)}q^{4}(\beta_{5}-\beta_{8}) & 4W^{(1,I)}_{4}q^{5}(\beta_{13}-\beta_{11})\\
  & &  &+2W_{4}^{(1,I)}q^{6}(\beta_{9}-\beta_{12}) &  &\\
& &  & &  &\\
 0& 0& 0 & W_{2}^{(1,I)}q^{3}(\beta_{4}-\beta_{6})  & 2W_{4}^{(1,I)}q^{3}(\beta_{4}-\beta_{6})& 4W^{(1,I)}_{4}q^{4}(\beta_{7}-\beta_{8})\\
  & & &+2W_{4}^{(1,I)}q^{5}(\beta_{11}-\beta_{13}) & & \\
 \end{array}\right)\nonumber\\
\end{eqnarray}
}

\noindent  Finally, the column matrix $B_{(0,I)}$ reads
$B_{(0,I)}=(\beta_{0},q^{2}\beta_{2},q\beta_{1},q^{4}\beta_{5},q^{3}\beta_{4},q^{2}\beta_{3},0,0,0).$

\subsection{Case S=1 M=0}

Similarly to the previous subsection, we can build a vector $X_{(1,0,I)}$ whose components are

\begin{tabular}{l l}
$X_{0} = \langle G_{RPA}^{(1,0,I)} \rangle$ & $X_{1} =  \langle k^{2}G_{RPA}^{(1,0,I)} \rangle$ \\
& \\
$X_{2} = \sqrt{\frac{4\pi}{3}}\langle k Y_{10}G_{RPA}^{(1,0,I)} \rangle$ & $X_{3} = \langle k^{4} G_{RPA}^{(1,0,I)} \rangle$\\
& \\
$X_{4} = \sqrt{\frac{4\pi}{3}} \langle k^{3} Y_{10}G_{RPA}^{(1,0,I)} \rangle$ & $X_{5} = \frac{4\pi}{3}\langle k^{2} Y_{10}^{2}G_{RPA}^{(1,0,I)} \rangle$\\
& \\
\end{tabular}

\hfill 

\noindent Since the spin-orbit does not contribute in this channel, a closed system of equations can be obtained with only six unknown quantities. The matrix $A_{(1,0,I)}$ can be deduced from $A_{(0,I)}$ by taking $C^{\nabla J}_{I}=0$ and by substituting $W_{1,2,4}^{(0,I)}$ by $W_{1,2,4}^{(1,I)}$.

\noindent The matrix $B_{(1,0,I)}$ reads
$B_{(1,0,I)}=(\beta_{0},q^{2}\beta_{2},q\beta_{1},q^{4}\beta_{5},q^{3}\beta_{4},q^{2}\beta_{3}).$

\subsection{Case S=1 M=1}

This channel is very close to $(0,I)$ ones. The vector of unknown quantities $X_{(1,1,I)}$ has the following components

\begin{tabular}{l l}
& \\
$X_{0}= \langle G_{RPA}^{(1,1,I)} \rangle$ & $X_{1} = \langle k^{2}G_{RPA}^{(1,1,I)} \rangle$\\
& \\
$X_{2}=\sqrt{\frac{4\pi}{3}}\langle k Y_{10}G_{RPA}^{(1,1,I)} \rangle$ & $X_{3}=\langle k^{4} G_{RPA}^{(1,1,I)} \rangle$\\
& \\
$X_{4}=\sqrt{\frac{4\pi}{3}} \langle k^{3} Y_{10}G_{RPA}^{(1,1,I)} \rangle$ & $X_{5}=\frac{4\pi}{3}\langle k^{2} Y_{10}^{2}G_{RPA}^{(1,1,I)} \rangle$\\
& \\
$Y_{1}=\sqrt{\frac{4\pi}{3}}\sum_{M'}M'\langle k Y_{1M'}G_{RPA}^{(0,I)}\rangle$ & $Y_{2}=\sqrt{\frac{4\pi}{3}}\sum_{M'} M' \langle k^{3} Y_{1M'}G^{(0,I)}_{RPA}\rangle$\\
& \\
$Y_{3}=\frac{4\pi}{3}\sum_{M'} M' \langle k^{2} Y_{1M'}Y_{10}G^{(0,I)}_{RPA}\rangle$ \\
& \\
\end{tabular}

\noindent The matrix $A_{(1,1,I)}$ can be deduced from $A_{(0,I)}$ by  simply substituting $W_{1,2,4}^{(0,I)}$ by $W_{1,2,4}^{(1,I)}$. Moreover, we have $B_{(0,I)} = B_{(1,1,I)}$.

\section{Expressions of $W^{(S,I)}_{i=1,4}$}
\label{app:cpl}

The $W^{(S,I)}_{i=1,4}$ coefficients expressions entering in the response functions in \ref{systems} can be expressed with respect to $4th$ order coupling constants of the Skyrme functional as indicated in the table below. Note that only the $4th$ order contribution is written:  $W^{(S,I)}_{1}$ also receive other contributions from the usual Skyrme functional that are not given here (see \cite{pas12} for explicit expressions).

\begin{table}[H]
\bc
\label{tab:Skyrme_int:2bodyEDF:fourthorder_C:coeff}
\caption{Coefficients of the normal part of the fourth order EDF, Eq.~(\ref{eq:fullEDF4}), 
as a function of the parameters of the pseudo-potential of Eq.~(\ref{eq:fullEDF4_even_np}). 
Missing entries are zero.}

\begin{tabular}{r cc cc cc cc cc cc cc cc cc}
                           &&                  &&                 &&                  &&                 \\[0.3mm] 
\hline \hline \noalign{\smallskip}
 && $C_0^{(4)\Delta\rho}$ && $C_1^{(4)\Delta\rho}$ && $C_0^{(4) M \rho}$ && $C_1^{(4) M \rho}$  &&  $C_0^{(4)\Delta s}$   &&    $C_1^{(4)\Delta s}$  &&    $C_0^{(4) M s}$  &&   $C_1^{(4) M s}$      \\
\noalign{\smallskip}  \hline \noalign{\smallskip}
$W^{(0,0)}_{1} $ && $8 q^4$ &&                 && $-\frac{1}{4}q^4$ &&   && && && && 
\\[0.3mm]
$W^{(0,1)}_{1} $ &&  && $ 8 q^4$ &&  && $-\frac{1}{4}q^4$  && && && && 
\\[0.3mm]
$W^{(1,0)}_{1} $    &&   &&         &&    &&    &&       $8 q^4$ &&                 && $-\frac{1}{4}q^4$ &&      \\[0.3mm]
$W^{(1,1)}_{1} $    &&   &&         &&    &&    &&       && $8 q^4$ &&                 && $-\frac{1}{4}q^4$ \\[0.3mm]
$W^{(0,0)}_{4} $    &&   &&         &&    $\frac{1}{2}$  && &&  &&  &&  && \\[0.3mm]
$W^{(0,1)}_{4} $    &&   &&         &&    && $\frac{1}{2}$ && &&  &&  &&             \\[0.3mm]
$W^{(1,0)}_{4} $    &&   &&         &&    &&     &&    &&  && $\frac{1}{2}$ && \\[0.3mm]
$W^{(1,1)}_{4} $    &&   &&         &&    &&     &&    &&  &&  &&  $\frac{1}{2}$         \\[0.3mm]
\noalign{\smallskip} \hline \hline
\end{tabular}
\ec
\end{table}


\end{appendix}




\section*{References}

\begin{thebibliography}{10}
\bibitem{RMP}
Bender, M., Heenen, P.-H., and Reinhard, P.-G.,
\newblock Rev. Mod. Phys. {\bf 75} (2003) 121.
\bibitem{Sky56a}
Skyrme, T. H.~R.,
\newblock Phil. Mag. {\bf 1} (1956) 1043.
\bibitem{Sky59b}
Skyrme, T. H.~R.,
\newblock Nucl. Phys. {\bf 9} (1958) 635.
\bibitem{Dec80a}
Decharg{\'e}, J. and Gogny, D.,
\newblock Phys. Rev. C {\bf 21} (1980) 1568.
\bibitem{VB72}
Vautherin, D. and Brink, D.~M.,
\newblock Phys. Rev. C {\bf 5} (1972) 626.
\bibitem{Rin80aB}
Ring, P. and Schuck, P.,
\newblock {\em The Nuclear Many Body Problem},
\newblock Springer, Berlin, 1980.
\bibitem{Cha97a}
Chabanat, E., Bonche, P., Haensel, P., Meyer, J., and Schaeffer, R.,
\newblock Nucl. Phys. {\bf A627} (1997) 710.
\bibitem{Cha98a}
Chabanat, E., Bonche, P., Haensel, P., Meyer, J., and Schaeffer, R.,
\newblock Nucl. Phys. {\bf A635} (1998) 231,
\newblock Erratum Nucl. Phys. {\bf A643}, 441 (1998).
\bibitem{per04}
Perli{\'n}ska, E., Rohozi{\'n}ski, S.~G., Dobaczewski, J., and Nazarewicz, W.,
\newblock Phys. Rev. C {\bf 69} (2004) 014316.
\bibitem{rai11}
Raimondi, F., Carlsson, B.~G., and Dobaczewski, J.,
\newblock Phys. Rev. C {\bf 83} (2011) 054311.

\bibitem{ber12}G. Bertsch, D. Dean, and W. Nazarewicz, SciDAC Review 6, 42 (2007); R. Furnstahl, Nucl. Phys. News 21,18 (2011); H. Nam, M. Stoitsov, W. Nazarewicz, A. Bulgac, G. Hagen, M. Kortelainen, P. Maris, J. C. Pei, K. J. Roche, N. Schunck, I. Thompson, J. P. Vary, and S. M.
Wild, J. Phys.: Conf. Ser. 402, 012033 (2012); R. Furnstahl, Nuclear Physics News \textbf{21}, 2, (2011)
\bibitem{bog13}S. Bogner \emph{et al.}, Comput. Phys. Comm. \textbf{184}, 2235 (2013).
\bibitem{link} http://computingnuclei.org.
\bibitem{kor10}
Kortelainen, M. et~al.,
\newblock Phys. Rev. C {\bf 82} (2010) 024313.
\bibitem{kor12}
Kortelainen, M. et~al.,
\newblock Phys. Rev. C {\bf 85} (2012) 024304.
\bibitem{kor14}
Kortelainen, M. et~al.,
\newblock Phys. Rev. C {\bf 89} (2014) 054314.
\bibitem{lac09}
Bender, M., Duguet, T., and Lacroix, D.,
\newblock Phys. Rev. C {\bf 79} (2009) 044319.
\bibitem{gor13}
Goriely, S., Chamel, N., and Pearson, J.~M.,
\newblock Phys. Rev. C {\bf 88} (2013) 061302.
\bibitem{Les07a}
Lesinski, T., Bender, M., Bennaceur, K., Duguet, T., and Meyer, J.,
\newblock Phys. Rev. C {\bf 76} (2007) 014312.
\bibitem{sad13}
Sadoudi, J., Duguet, T., Meyer, J., and Bender, M.,
\newblock Phys. Rev. C submitted.
\bibitem{car08}
Carlsson, B.~G., Dobaczewski, J., and Kortelainen, M.,
\newblock Phys. Rev. C {\bf 78} (2008) 044326.

\bibitem{car11}
Carlsson, B.~G. and Dobaczewski, J.,
\newblock Phys. Rev. Lett. {\bf 105} (2011) 122501.
\bibitem{dav13}
Davesne, D., Pastore, A., and Navarro, J.,
\newblock J. Phys. G: Nucl. Part. Phys. {\bf 40} (2013)
  095104.
\bibitem{dob96}
Dobaczewski, J. and Dudek, J.,
\newblock Acta Phys. Pol. {\bf B27} (1996) 45.
\bibitem{rai11b}
Raimondi, F., Carlsson, B.~G., Dobaczewski, J., and Toivanen, J.,
\newblock Phys. Rev. C {\bf 84} (2011) 064303.
\bibitem{dav14b}
Davesne, D., Pastore, A., and Navarro, J.,
\newblock J. Phys. G: Nucl. Part. Phys. {\bf 41} (2014)
  065104.
\bibitem{Sla51}
Slater, J.~C.,
\newblock Phys. Rev. {\bf 81} (1951) 385.
\bibitem{Lacroix2}
Lacroix, D., Duguet, T., and Bender, M.,
\newblock Phys. Rev. C {\bf 79} (2009) 044318.
\bibitem{Lacroix3}
Duguet, T., Bender, M., Bennaceur, K., Lacroix, D., and Lesinski, T.,
\newblock Phys. Rev. C {\bf 79} (2009) 044320.
\bibitem{cha10}
Chamel, N.,
\newblock Phys. Rev. C {\bf 82} (2010) 061307.
\bibitem{was12}
K. Washiyama, K. Bennaceur, B. Avez, M. Bender, P.-H. Heenen, and V. Hellemans
Phys. Rev. C \textbf{86}, 054309
\bibitem{Cha08a}
Chamel, N., Goriely, S., and Pearson, J.~M.,
\newblock Nucl. Phys. A {\bf 812} (2008) 72.

\bibitem{Dob00a}
Dobaczewski, J., Dudek, J., Rohozi{\'n}ski, S.~G., and Werner, T.~R.,
\newblock Phys. Rev. C {\bf 62} (2000) 014310.
\bibitem{Bon87a}
Bonche, P., Flocard, H., and Heenen, P.~H.,
\newblock Nucl. Phys. {\bf A467} (1987) 115.
\bibitem{gar92}
C. Garc\'{i}a-Recio and J. Navarro and Nguyen {Van Giai}
              and Salcedo, L. L.,
\newblock Ann. Phys. (N.-Y.) {\bf 214} (1992) 214.
\bibitem{Dav09a}
Davesne, D., Martini, M., Bennaceur, K., and Meyer, J.,
\newblock Phys. Rev. C {\bf 80} (2009) 024314.

\bibitem{pas12}
A. Pastore, D. Davesne, Y. Lallouet, M. Martini, K. Bennaceur, and J. Meyer, \newblock Phys. Rev. C {\bf 85} (2012) 054317.

\bibitem{pas12b}
A. Pastore, M. Martini, V. Buridon, D. Davesne, K. Bennaceur, and J. Meyer,
\newblock Phys. Rev. C {\bf 86} (2012) 044308.

\bibitem{Les06a}
Lesinski, T., Bennaceur, K., Duguet, T., and Meyer, J.,
\newblock Phys. Rev. C {\bf 74} (2006) 044315.
\bibitem{pas12c}
Pastore, A., Bennaceur, K., Davesne, D., and Meyer, J.,
\newblock J. Mod. Phys. E {\bf 5} (2012) 1250041.
\bibitem{hel12}
Hellemans, V., Heenen, P.-H., and Bender, M.,
\newblock Phys. Rev. C {\bf 85} (2012) 014326.
\bibitem{sch10}
Schunck, N. et~al.,
\newblock Phys. Rev. C {\bf 81} (2010) 024316.
\bibitem{hel13}
Hellemans, V. et~al.,
\newblock Phys. Rev. C {\bf 88} (2013) 064323.
\bibitem{pas13c}
Pastore, A., Davesne, D., Bennaceur, K., Meyer, J., and Hellemans, V.,
\newblock Physica Scripta {\bf 2013} (2013) 014014.




\bibitem{fet71}
Fetter, A.~L. and Walecka, J.~D.,
\newblock {\em Quantum Theory of Many-Particle Systems},
\newblock McGraw-Hill, New York, 1971.
\bibitem{heb09}
Hebeler, K., Duguet, T., Lesinski, T., and Schwenk, A.,
\newblock Phys. Rev. C {\bf 80} (2009) 044321.

\bibitem{kor10R}M. Kortelainen, R. J. Furnstahl, W. Nazarewicz, and M. V. Stoitsov, Phys. Rev. C \textbf{82}, 011304(R) (2010).

\bibitem{pas14}
Pastore, A., Davesne, D., and Navarro, J.,
\newblock J.Phys. G: Nucl. Part. Phys. {\bf G41} (2014) 055103.

\bibitem{boh79}
Bohigas, O. and Lane, A.M. and Martorell, J.,
\newblock   Phys. Rep. {\bf 51} (1979) 267.


\end{thebibliography}
\end{document}